\documentclass[preprint,3p,authoryear]{elsarticle}
\usepackage{graphicx}
\usepackage{amssymb}
\usepackage{lineno}
\usepackage{tabularx}
\begin{document}
\begin{frontmatter}
\title{PROPAGATION OF ACOUSTIC-GRAVITY WAVES IN NON-UNIFORM WIND FLOWS OF THE POLAR ATMOSPHERE}
\author{A. K. Fedorenko}
\ead{fedorenkoak@gmail.com}
\author{E. I. Kryuchkov}
\ead{kryuchkov.ye@gmail.com}
\author{O. K. Cheremnykh}
\ead{oleg.cheremnykh@gmail.com}
\author{Yu. O. Klymenko$^*$}
\ead{yurklym@gmail.com}
\cortext[cor1]{Corresponding author.}
\author{S.V. Melnychuk}
\ead{melnychuk89s@gmail.com}

\address{Space Research Institute of  the National Academy of Sciences of Ukraine and the State Space Agency of Ukraine, prosp. Akad. Glushkova 40, build. 4/1, 03187, Kyiv, Ukraine}

\begin{abstract}
Satellite observations of acoustic-gravity waves in the polar regions of the atmosphere indicate a close connection of these waves with wind flows. The paper investigates the peculiarities of the propagation of AGWs in spatially inhomogeneous flows, where the wind speed slowly changes in the horizontal direction. A system of hydrodynamic equations is used for the analysis, which takes into account the wind flow with spatial inhomogeneity. Unlike the system of equations written for a stationary medium (or a medium moving at a uniform speed), the resulting system contains components that describe the interaction of the waves with the medium. It is shown that the influence of inhomogeneous background parameters of the medium can be separated from the effects of inertial forces by means of the special variable substitution. The analytical expression is obtained that describes the changes in the wave amplitude in a medium moving with a non-uniform speed. This expression contains two functional dependencies: 1) the linear part caused by changes in the background parameters of the medium, it does not depend on the direction of the wave propagation relative to the flow; 2) the exponential part associated with the inertia forces, which determines the dependence of the amplitudes of AGWs on the direction of their propagation. The exponential part shows an increase in the amplitudes of the waves in the headwind and a decrease in their amplitudes in the downwind. The obtained theoretical dependence of the amplitudes of AGWs on the wind speed is in good agreement with the data of satellite observations of these waves in the polar thermosphere.

\end{abstract}
\end{frontmatter}

\section{INTRODUCTION}

The polar thermosphere is the important region of the atmosphere where there is a close interaction of various dynamic processes, including waves with wind flows. Namely, in the high latitudes, there are the most powerful sources of atmospheric disturbances, in particually, various manifestations of geomagnetic activity. In addition, an unique wind system is formed in these areas, which is the result of the superposition of the planetary wind circulation and the drift movements of the ionospheric plasma. In contrast to middle and low latitudes, the spatial structure of the polar winds depends to a lesser extent on the conditions of illumination by the Sun than on the processes of magnetospheric-ionospheric interaction. These processes lead to the formation of vortices in the polar upper atmosphere with very high flow velocities ranging from $\sim$250 m/s to $\sim$700 m/s depending on geomagnetic activity (Killeen et al.,1995; L$\rm\ddot{u}$hr et al., 2007).
	
An analysis of measurements of acoustic-gravity wave (AGW) parameters on the Dynamics Explorer 2 satellite indicates a close connection of atmospheric wave disturbances in the polar thermosphere with the wind circulation. According to these data, the acoustic-gravity waves of large amplitudes are regularly observed in the regions of powerful polar wind systems (Innis, Conde, 2002; Fedorenko et al., 2015). The observed waves propagate towards the wind, and their amplitude is approximately proportional to the wind speed (Fedorenko et al., 2015). It is logical to assume that the features of AGWs observed in the polar regions are the result of their propagation in spatially inhomogeneous wind flows.

The main features of AGWs in non-uniform atmospheric flows are the filtering of the wave spectrum and the change in their amplitudes as a result of interaction with the medium. The filtering of the AGW spectrum in a spatially inhomogeneous flow, the speed of which depends on the spatial coordinates, was previously investigated in (Fedorenko et al., 2023). It was shown that in an headwind, the direction and magnitude of the wave vector change as follows: with increasing wind speed, the wave vector gradually inclines to the horizontal plane, and its horizontal component tends to some limiting value. The internal frequencies of the waves increase in the medium, approaching the Brunt-V\"{a}is\"{a}l\"{a} frequency in sufficiently strong winds. These effects are actually manifestations of AGW filtering, as a result of which from the continuous spectrum of atmospheric waves, generating by a hypothetical source, it is remained only high-frequency harmonics with a small angle of inclination of the wave vector to the horizontal plane and a certain characteristic wavelength in a strong headwind. Note that satellite observations indicate a predominance of waves with such characteristics (Innis, Conde, 2002; Fedorenko et al., 2015).

Changes in the amplitudes of AGW in a non-uniform wind flow were previously studied by us in several works. The easiest way to estimate the amplitude of the AGW in a spatially inhomoogeneous atmospheric flow is by using the wave action equation (Bretherton, Garrett, 1969; Fedorenko et al., 2018). The work (Fedorenko et al., 2022) analyzed the change in the amplitudes of AGW in the flow based on the system of linearized hydrodynamic equations written for the vertical and the horizontal components of displacements of the elementary gas volume. In order to take into account the interaction of waves with the flow, inertial forces were added to the first equation for the horizontal displacement component in (Fedorenko et al., 2022) following (Lighthill, 1978). In the framework of this approach, an analytical expression was obtained that approximately describes the changes in the wave amplitudes in an inhomogeneous flow and it is in good agreement with experimental data. However, our more detailed analysis using 4 first-order hydrodynamic equations for velocity vector, density, and pressure disturbances showed that in paper (Fedorenko et al., 2022) not all components responsible for the interaction of the waves with the medium were taken into account. Therefore, in this work, we took into account these new components and analyzed how they affect the resulting changes in the amplitude of waves in the flow.

We obtaine an analytical expression that describes the changes in the amplitude of waves in a slightly inhomogeneous moving medium. When it is obtained, the inertial forces and the changes in the background density of the atmosphere in a non-uniform flow are taken into account. It is shown that these effects can be separated from each other by a special substitution of variables. The influence of the inertial forces on the AGW amplitude depends on the direction of the wave propagation relative to the wind flow. Namely, the waves are attenuated in the fair wind and increased in the headwind.

\section{INTERACTION OF AGW WITH WIND ACCORDING TO SATELLITE DATA}

Satellite observations indicate the predominance of AGWs with certain spectral properties and selected propagation azimuths in the polar thermosphere. So, the frequencies of AGW in the polar thermosphere, determined from the satellite measurements, are mostly close to the Brunt-V\"{a}is\"{a}l\"{a} frequency, and their horizontal scales are approximately 500 km (Innis, Conde, 2002; Fedorenko et al., 2015). In the presence of various hypothetical sources in the polar atmosphere, these experimental results probably indicate selective filtering of the AGW spectrum. From the analysis of the measurements on the Dynamics Explorer 2 (DE2) satellite, it was also established that the behavior of AGW in the polar regions is mainly controlled by winds (Fedorenko et al., 2015). This wind control mainly manifests itself in two effects: 1) systematic AGWs propagating towards the wind; 2) dependence of wave amplitudes on wind speed.

The dependence of the amplitude of polar AGWs on the speed of the headwind is shown in Fig.1. AGWs are considered in the fluctuations of the atmosphere density $\Delta \rho /\rho $ (Fig. 1a) and in the vertical component of the particle velocity $v_{z} $  (Fig. 1b). Data of measurements of the neutral atmosphere parameters on the low-orbit polar satellite Dynamics Explorer 2 are used. The Figures show the relative change in the amplitude $A/A_{0} $  of the waves. Here, the normalization is made on the amplitude of some background level of AGW, which corresponds to the background wind with the speed  $W_{x} \approx 50$m/s.

Concentrations of different atmospheric gases are used to calculate the density fluctuations, which were measured on the DE2 satellite using a mass spectrometer in NACS (Neutral Atmosphere Composition Spectrometer) experiment (Carignan et al., 1981). The temperature and vertical velocity of neutral particles were measured in WATS (Wind and Temperature Spectrometer) experiment (Spencer et al., 1981). As can be seen from Fig. 1, the angles of inclination of the interpolation lines for the wind dependence of the density and velocity amplitude fluctuations practically coincide. It confirms the consistent increase in the amplitude of AGW in various atmospheric parameters when they propagate in the headwind. It should be noted that the concentration and the velocity measurements are carried out with different devices, while significant background differences in the atmospheric density were excluded in order to obtain the density fluctuations. Therefore, the agreement of these experimental dependencies indicates both the high accuracy of the measurements and the correctness of the data processing procedure.
\begin{figure}
\centering
\includegraphics[scale=0.55]{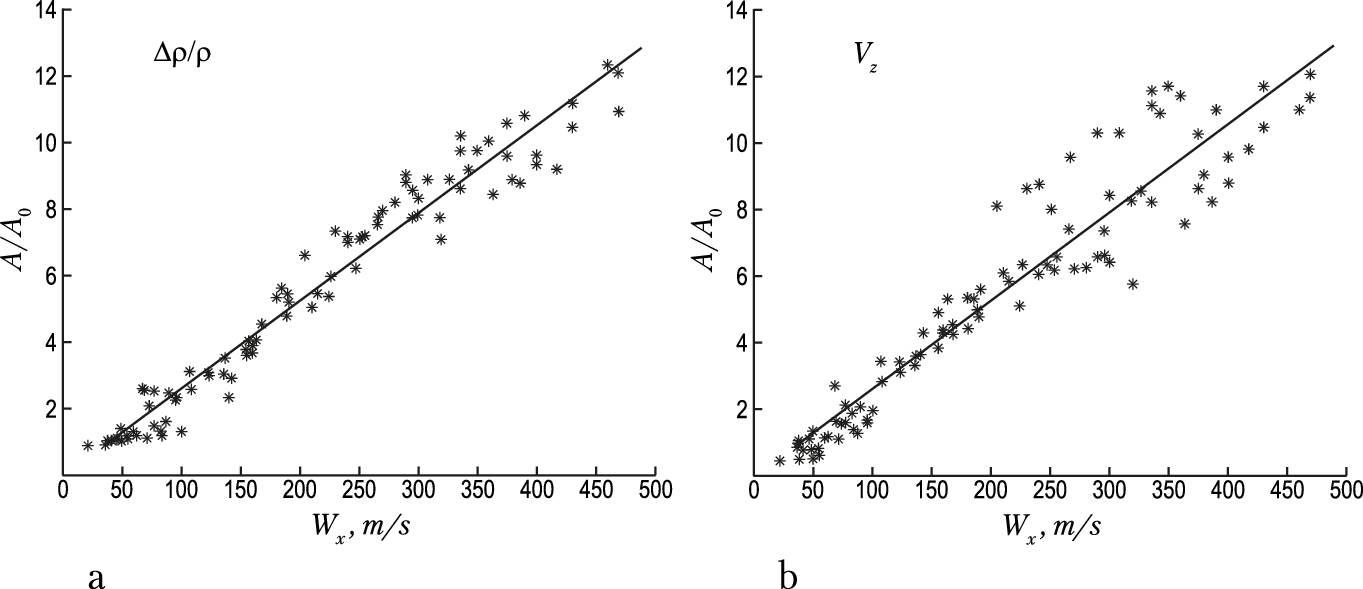}
\caption{Dependence of AGW amplitude on the speed of the headwind in the polar regions according to the Dynamics Explorer 2 satellite measurements: a) $\Delta \rho /\rho $ is density fluctuations according to NACS experiment, $A/A_{0} =0.0263W_{x} -0.0523$; b) $V_{z} $ is vertical velocity fluctuations according to WATS experiment, $A/A_{0} =0.0265W_{x} -0.0594$.}
\end{figure}

\section{MAIN EQUATIONS }

In order to explain the results of the satellite observations discussed in the previous section, let's investigate the propagation of AGW in a horizontally inhomogeneous wind flow. It is known that the wind flows in the polar thermosphere are formed as a result of the interaction of neutral atmosphere with ionospheric plasma drifting under the influence of magnetic and electric fields (Killeen et al., 1995; L$\rm\ddot{u}$hr et al., 2007). The spatial configuration of the polar flows in the thermosphere mainly depends on geomagnetic activity. We will not be interested in the physical nature of the global inhomogeneous flow, against which waves propagate. We consider this wind flow as given, and assume that its influence on the AGW is realized through changes in the background parameters of the medium and the inertia force, which describe the interaction of waves with the medium.

Let us assume that the flow velocity is directed along the horizontal coordinate $x$ and it slowly changes in the direction of propagation: $\vec{W}=\left(W_{x} ,0,0\right)$ and $\partial W_{x} /\partial x\ne 0$. Assumption $\partial W_{x} /\partial z=0$ is physically justified for the heights of the thermosphere, since the wind speed in the atmosphere above $\sim$200 km changes slowly with the height due to significant influence of molecular viscosity (Vadas, Fritts, 2005).

To consider the wave disturbances in a non-uniform atmospheric flow, we use the well-known system of hydrodynamic equations, which includes equations of the motion, continuity, and heat conservation. In the atmosphere stratified in the gravity field and in the absence of external forces, mass and heat inflow, the system of equations has the form:
\begin{equation} \label{1}
\rho \frac{DV_{x} }{Dt} +\frac{\partial P}{\partial x} =0,
\end{equation}
\begin{equation} \label{2}
\rho \frac{DV_{z} }{Dt} +\frac{\partial P}{\partial z} +\rho g=0
\end{equation}
\begin{equation} \label{3}
\frac{D\rho }{Dt} +\rho \nabla \vec{V}=0,
\end{equation}
\begin{equation} \label{4}
\frac{DP}{Dt} =\gamma \frac{P}{\rho } \frac{D\rho }{Dt}.
\end{equation}
Here, $\rho $ is the atmosphere density, $P$ is the pressure, $\vec{V}\left(V_{x} ,V_{z} \right)$ is the velocity vector of the atmospheric medium, $g$ is the acceleration of the gravity, $\gamma $ is the special heats ratio, and $\frac{D}{Dt} =\frac{\partial }{\partial t} +\left(\vec{V}\cdot \nabla \right)$ is the substantial derivative.

We will consider an isothermal atmosphere in which the undisturbed density and the undisturbed pressure satisfy the condition of hydrostatic equilibrium:
\[\frac{1}{P_{0} } \frac{\partial P_{0} }{\partial z} =\frac{1}{\rho _{0} } \frac{\partial \rho _{0} }{\partial z} =-\frac{1}{H} ,\]
where $H$ is the atmosphere scale height. For a stationary atmosphere, we can write $V_{x} =v_{x} $, $V_{z} =v_{z} $, $\rho =\rho _{0} +\rho '$, and $P=P_{0} +P'$, where $\rho _{0} $, $P_{0} $ are the undisturbed density and the pressure, $\rho '$ and $P'$ are the density and the pressure perturbations, $v_{x} $ and $v_{z} $ are the velocity perturbations in the vertical and the horizontal directions. Then, for small atmospheric disturbances, from (\ref{1}) -- (\ref{4}) it follows the well-known linearized system of hydrodynamic equations (Hines, 1969; Nappo, 2002):
\begin{equation} \label{5}
\rho _{0} \frac{\partial v_{x} }{\partial t} +\frac{\partial P'}{\partial x} =0,
\end{equation}
\begin{equation} \label{6}
\rho _{0} \frac{\partial v_{z} }{\partial t} +\frac{\partial P'}{\partial z} +\rho 'g=0,
\end{equation}
\begin{equation} \label{7}
\frac{\partial \rho '}{\partial t} +\frac{\partial \rho _{0} }{\partial z} v_{z} +\rho _{0} \nabla \vec{v}=0,
\end{equation}
\begin{equation} \label{8}
\frac{\partial P'}{\partial t} +\frac{\partial P_{0} }{\partial z} v_{z} +\gamma P_{0} \nabla \vec{v}=0.
\end{equation}
System (\ref{5}) -- (\ref{8}) is often used to analyze atmospheric AGWs. From this system it is possible to obtain the main relations of the linear theory of AGW for an isothermal atmosphere: the dispersion equation and the polarization relations between the fluctuations of various perturbed quantities.

If the medium moves horizontally with a certain speed $W_{x} $, then $V_{x} =W_{x} +v_{x} $. In this case, in addition to partial time derivatives, the substantial derivatives at $W_{x} =const$ for various quantities  contain the terms with spatial derivatives:
\begin{equation} \label{9}
\frac{D}{Dt} =\frac{\partial }{\partial t} +W_{x} \frac{\partial }{\partial x} +\nu _{x} \frac{\partial }{\partial x} +\nu _{z} \frac{\partial }{\partial z}.
\end{equation}
The first two terms, after the substitution in (\ref{9}) the solution in the form of monochromatic wave $v_{x} ,v_{z} ,\rho ',P'\propto \exp \left[i(\omega _{0} t-k_{x} x-k_{z} z)\right]$, lead to frequency renormalization $\omega =\omega _{0} -k_{x} W_{x} $. Here, $\omega _{0} $ is the frequency in the stationary medium, $k_{x} $ and $k_{z} $ are the horizontal and the vertical components of the wave vector. Therefore, at $W_{x} =const$ we can go to the moving reference medium, in which the form of the differential hydrodynamic equations (\ref{5})--(\ref{8}) for the internal frequency $\omega $ remains the same as for the stationary medium.

If the wind speed $W_{x} \left(x\right)$ depends on the horizontal coordinate, but it changes slowly on the wavelength scale, it is also possible to go to the reference frame of the medium (Lighthill, 1978). At the same time, $\omega $ and $k_{x} $ are considered as local constants. However, the system of hydrodynamic equations (\ref{5})--(\ref{8}) contain a number of new terms in this case. Taking  into account (\ref{9}), it will take the following form:
\begin{equation} \label{10}
\rho _{0} \frac{\partial v_{x} }{\partial t} +\frac{\partial P'}{\partial x} =-\rho _{0} v_{x} \frac{\partial W_{x} }{\partial x} -\rho 'W_{x} \frac{\partial W_{x} }{\partial x},
\end{equation}
\begin{equation} \label{11}
\rho _{0} \frac{\partial v_{z} }{\partial t} +\frac{\partial P'}{\partial z} +\rho 'g=0,
\end{equation}
\begin{equation} \label{12}
\frac{\partial \rho '}{\partial t} +\frac{\partial \rho _{0} }{\partial z} v_{z} +\frac{\partial \rho _{0} }{\partial x} v_{x} +\rho _{0} \nabla \vec{v}=-\rho '\frac{\partial W_{x} }{\partial x},
\end{equation}
\begin{equation} \label{13}
\frac{\partial P'}{\partial t} +\frac{\partial P_{0} }{\partial z} v_{z} +\frac{\partial P_{0} }{\partial x} v_{x} +\gamma P_{0} \nabla \vec{v}=-\gamma P'\frac{\partial W_{x} }{\partial x}.
\end{equation}
Here, the new terms in the right-hand side of equations (\ref{10}), (\ref{12}), and (\ref{13}), which are proportional to the gradient of the flow velocity $\partial W_{x} /\partial x$, are the inertia forces. They are responsible for the interaction of waves with the moving medium. Also, in Eqs. (\ref{12}) and (\ref{13}) there are new terms that take into account changes in the background parameters of the medium along the horizontal coordinate $\partial \rho _{0} /\partial x$ and $\partial P_{0} /\partial x$. We also suppose that the condition of the hydrostatic equilibrium along the vertical coordinate is fulfilled if the the medium velocity changes slowly.

We considered the system of the first order equations in order to correctly take into account all components arising in an inhomogeneous flow due to the time differentiation. For  convenience, let's go from four equations (\ref{10})--(\ref{13}) to two second-order equations written for fluctuations of the velocity components $v_{x} $ and $v_{z} $. As will be shown below, this allows to make some substitution of variables simplifying the analytical consideration of the equations. For this purpose, we differentiate Eq. (\ref{10}) by time $t$ and Eq. (\ref{13}) by horizontal coordinate $x$. Then, we subtract one equation from the other, thereby eliminating from the variable $P'$, and obtain equation for variable $v_{x} $. After that, we differentiate Eq. (\ref{11}) by $t$, and Eq. (\ref{13}) by $z$, and again substract one equation with another. By substituting the values of $\partial \rho '/\partial t$ from (\ref{12}), we can get the equation for $v_{z} $. Finally, we have the following system for the velocity components:
\begin{equation} \label{14}
\rho _{0} \frac{\partial ^{2} v_{x} }{\partial t^{2} } -\frac{\partial }{\partial x} \left(\lambda e\right)+g\frac{\partial }{\partial x} \left(\rho _{0} v_{z} \right)-c_{s}^{2} \frac{\partial }{\partial x} \left(\frac{\partial \rho _{0} }{\partial x} v_{x} \right)=F_{i1},
\end{equation}
\begin{equation} \label{15}
\rho _{0} \frac{\partial ^{2} v_{z} }{\partial t^{2} } -\frac{\partial }{\partial z} \left(\lambda e\right)-g\frac{\partial }{\partial x} \left(\rho _{0} v_{x} \right)-c_{s}^{2} \frac{\partial }{\partial z} \left(\frac{\partial \rho _{0} }{\partial x} v_{x} \right)=F_{i2},
\end{equation}
where $\lambda =\rho _{0} c_{s}^{2} $; $e=\frac{\partial v_{x} }{\partial x} +\frac{\partial v_{z} }{\partial z} $, $c_{s} $ is the sound velocity, $F_{i1} =-\rho _{0} \left[\left(1+\gamma \right)\frac{\partial v_{x} }{\partial t} +\frac{W_{x} }{\rho _{0} } \frac{\partial \rho '}{\partial t} \right]\frac{\partial W_{x} }{\partial x} $ , $F_{i2} =-\rho _{0} \gamma \left[\frac{\partial v_{z} }{\partial t} +N^{2} H\left(\frac{\rho '}{\rho _{0} } \right)\right]\frac{\partial W_{x} }{\partial x} $, $N^{2} =\frac{g}{H} \frac{\gamma -1}{\gamma } $  is the square of the Brunt-V\"{a}is\"{a}l\"{a} frequency (BV).

\section{SUBSTITUTION OF VARIABLES}

Based on the continuity of the background flow we get $\frac{1}{\rho _{0} } \frac{\partial \rho _{0} }{\partial x} =-\frac{1}{W_{x} } \frac{\partial W_{x} }{\partial x} $ and Eqs. (\ref{14}), (\ref{15}) can be rewritten in the form:
\begin{equation} \label{16}
\rho _{0} \frac{\partial ^{2} }{\partial t^{2} } \left(\frac{v_{x} }{W_{x} } \right)-\lambda \frac{\partial }{\partial x} \left[\frac{\partial }{\partial x} \left(\frac{v_{x} }{W_{x} } \right)+\frac{\partial }{\partial z} \left(\frac{v_{z} }{W_{x} } \right)\right]+\rho _{0} g\frac{\partial }{\partial x} \left(\frac{v_{z} }{W_{x} } \right)=\frac{F_{i1} }{W_{x} },
\end{equation}
\begin{equation} \label{17}
\rho _{0} \frac{\partial ^{2} }{\partial t^{2} } \left(\frac{v_{z} }{W_{x} } \right)-\frac{\partial }{\partial z} \left[\lambda \left(\frac{\partial }{\partial x} \left(\frac{v_{x} }{W_{x} } \right)+\frac{\partial }{\partial z} \left(\frac{v_{z} }{W_{x} } \right)\right)\right]-\rho _{0} g\frac{\partial }{\partial x} \left(\frac{v_{x} }{W_{x} } \right)=\frac{F_{i2} }{W_{x} }.
\end{equation}
From these equations, it follows the appropriateness of entering new variables $\tilde{v}_{x} =v_{x} /W_{x} $ and $\tilde{v}_{z} =v_{z} /W_{x} $. As a result of this substitutions, we obtain:
\begin{equation} \label{18}
\rho _{0} \frac{\partial ^{2} \tilde{v}_{x} }{\partial t^{2} } -\lambda \frac{\partial }{\partial x} \left[\frac{\partial \tilde{v}_{x} }{\partial x} +\frac{\partial \tilde{v}_{z} }{\partial z} \right]+\rho _{0} g\frac{\partial \tilde{v}_{z} }{\partial x} =\tilde{F}_{i1},
\end{equation}
\begin{equation} \label{19}
\rho _{0} \frac{\partial ^{2} \tilde{v}_{z} }{\partial t^{2} } -\frac{\partial }{\partial z} \left[\lambda \left(\frac{\partial \tilde{v}_{x} }{\partial x} +\frac{\partial \tilde{v}_{z} }{\partial z} \right)\right]-\rho _{0} g\frac{\partial \tilde{v}_{x} }{\partial x} =\tilde{F}_{i2}.
\end{equation}
where $\tilde{F}_{i1} =F_{i1} /W_{x} $ and $\tilde{F}_{i2} =F_{i2} /W_{x} $.

For comparison, we consider the similar system of equations obtained in (Tolstoy, 1963) for a stationary medium:
\begin{equation} \label{20}
\rho _{0} \frac{\partial ^{2} v_{x} }{\partial t^{2} } -\frac{\partial }{\partial x} \left(\lambda e\right)+\rho _{0} g\frac{\partial v_{z} }{\partial x} =0,
\end{equation}
\begin{equation} \label{21}
\rho _{0} \frac{\partial ^{2} v_{z} }{\partial t^{2} } -\frac{\partial }{\partial z} \left(\lambda e\right)-\rho _{0} g\frac{\partial v_{x} }{\partial x} =0.
\end{equation}
For a horizontally homogeneous medium, the value of  $\lambda $ in expression (\ref{20}) can be removed from the derivative sign. Therefore, the left-hand parts of equations (\ref{18}), (\ref{19}) in the new variables $\tilde{v}_{x} $, $\tilde{v}_{z} $ coincide with the left-hand parts of equations (\ref{20}), (\ref{21}) for usual variables $v_{x} $ and $v_{z} $. The difference between these equations remained only in the right-hand side (the inertia forces). In fact, with the help of substitutions $\tilde{v}_{x} =v_{x} /W_{x} $, $\tilde{v}_{z} =v_{z} /W_{x} $, the change in the background parameters of the medium has been taken into account. This indicates that there is a linear part of the dependence of the wave amplitude on the flow speed, which does not depend on the direction of their propagation. As will be shown below, the dependence of the wave amplitude on the wind direction is determined by the inertial forces.

\section{DISPERSION EQUATION AND DEPENDENCE OF AGW AMPLITUDE ON WIND SPEED}

For an isothermal stratified atmosphere, we look for the solution of Eqs. (\ref{18}), (\ref{19}) in the form
\begin{equation} \label{22}
\tilde{v}_{x} ,\tilde{v}_{z} \propto \exp \left(z/2H\right)\exp \left[i(\omega t-k_{x} x-k_{z} z)\right].
\end{equation}
Here, the dependence of the amplitude on the height is given taking into account the barometric stratification of the atmosphere (Hines, 1969; Nappo, 2002).

In the expressions for the inertial forces, we express the fluctuations of $\rho '$ in terms of the velocity components using the system of equations (\ref{10}) -- (\ref{13}):
\[\frac{\partial }{\partial t} \left(\frac{\rho '}{\rho _{0} } \right)=-W_{x} \left[\frac{\partial \tilde{v}_{x} }{\partial x} +\frac{\partial \tilde{v}_{z} }{\partial z} -\frac{\tilde{v}_{z} }{H} \right].\]
Using this expression, we rewrite the inertial forces in the new variables as follows:
\[\tilde{F}_{i1} =-\rho _{0} \frac{\partial W_{x} }{\partial x} \left[i\omega \left(1+\gamma \right)\tilde{v}_{x} -W_{x} \left(\frac{\partial \tilde{v}_{x} }{\partial x} +\frac{\partial \tilde{v}_{z} }{\partial z} -\frac{\tilde{v}_{z} }{H} \right)\right], \tilde{F}_{i2} =-i\gamma \omega \rho _{0} \frac{\partial W_{x} }{\partial x} \left[\tilde{v}_{z} -\frac{N^{2} H}{\omega ^{2} } \left(\frac{\partial \tilde{v}_{x} }{\partial x} +\frac{\partial \tilde{v}_{z} }{\partial z} -\frac{\tilde{v}_{z} }{H} \right)\right].\]
After substituting the solution (\ref{22}) into equations (\ref{18}), (\ref{19}) and under condition that the determinant of the system is equal to zero, we obtain the following relation:
\[
\omega ^{4} -\omega ^{2} c_{s}^{2} \left(\frac{1}{4H^{2} } +k_{x}^{2} +k_{z}^{2} \right)+k_{x}^{2} N^{2} c_{s}^{2} +i\gamma \omega \left(k_{x}^{2} c_{s}^{2} -\omega ^{2} \right)\frac{\partial W_{x} }{\partial x} +
\]
\begin{equation} \label{23}
+i\left(\gamma +1\right)\omega \left(\frac{c_{s}^{2} }{4H^{2} } +k_{z}^{2} c_{s}^{2} -\omega ^{2} \right)\frac{\partial W_{x} }{\partial x}
-\gamma Hk_{z} N^{2} \left(\omega +k_{x} W_{x} \right)\frac{\partial W_{x} }{\partial x}+
\end{equation}
\[
+ik_{x} W_{x} \left(\frac{\gamma }{2} N^{2} -\omega ^{2} \right)\frac{\partial W_{x} }{\partial x} +i\omega N^{2} \left(\frac{\gamma }{2} -\left(\gamma -1\right)\frac{k_{x}^{2} c_{s}^{2} }{\omega ^{2} } \right)\frac{\partial W_{x} }{\partial x} =0.
\]
The complexity of equation (\ref{23}) indicates a change in the amplitude of the waves. Let's determine the decrement of the change in the amplitude of the AGW in the wind flow. To do this, we will make the following substitution $k_{x} =\tilde{k}_{x} +i\delta $ in Eq. (\ref{23}), where $\delta $ is the spatial decrement of the amplitude change. As a result, we obtain the following equation from Eq. (\ref{23}) with accuracy up to linear terms in small parameters $\delta $ and $\partial W_{x} /\partial x$:
\[
\omega ^{4} -\omega ^{2} c_{s}^{2} \left(\frac{1}{4H^{2} } +\tilde{k}_{x}^{2} +k_{z}^{2} \right)+\tilde{k}_{x}^{2} N^{2} c_{s}^{2} +2i\delta \tilde{k}_{x} c_{s}^{2} \left(N^{2} -\omega ^{2} \right)-
\]
\begin{equation} \label{24}
-\gamma Hk_{z} N^{2} \left(\omega +\tilde{k}_{x} W_{x} \right)\frac{\partial W_{x} }{\partial x} +
+i\omega \left[\gamma \left(\tilde{k}_{x}^{2} c_{s}^{2} -\omega ^{2} \right)+\left(\gamma +1\right)\left(\frac{c_{s}^{2} }{4H^{2} } +k_{z}^{2} c_{s}^{2} -\omega ^{2} \right)\right]\frac{\partial W_{x} }{\partial x}+
\end{equation}
\[
+i\omega \left[\frac{W_{x} \tilde{k}_{x} }{\omega } \left(\frac{\gamma }{2} N^{2} -\omega ^{2} \right)+N^{2} \left(\frac{\gamma }{2} -\left(\gamma -1\right)\frac{\tilde{k}_{x}^{2} c_{s}^{2} }{\omega ^{2} } \right)\right]\frac{\partial W_{x} }{\partial x} =0.
\]
The real part of expression (\ref{24}) is the dispersion equation of AGW in a non-uniform flow, it coincides with the dispersion equation for a stationary medium, except for one term $\gamma Hk_{z} N^{2} \left(\omega +\tilde{k}_{x} W_{x} \right)\frac{\partial W_{x} }{\partial x} $. This term is small with a slow change of the flow velocity.

Equating the imaginary part of Eq. (\ref{24}) to zero and after some algebraic transformations, we obtain the expression for the spatial decrement of the change in the AGW amplitude for the inhomogeneous flow:
\begin{equation}\label{25} \delta =-\frac{1}{2} \frac{\partial W_{x} }{\partial x} \left[\frac{1}{U_{x} } \frac{\left(2\tau ^{2} -1\right)}{\left(\tau ^{2} -1\right)} +\frac{W_{x} }{2c_{s}^{2} } \frac{\left(\gamma \tau ^{2} -2\right)}{\left(\tau ^{2} -1\right)} +\frac{\gamma U_{x} }{2c_{s}^{2} } \frac{\left(\tau ^{2} -2\right)}{\left(\tau ^{2} -1\right)} \right].
\end{equation}
Here, $U_{x} =\omega /\tilde{k}_{x} $ is horizontal phase speed of the wave and $\tau ^{2} =N^{2} /\omega ^{2} $. When obtaining Eq. (\ref{25}), it was used that the well-known dispersion equation of AGW is approximately satisfied
\[\omega ^{4} -\omega ^{2} c_{s}^{2} \left(\frac{1}{4H^{2} } +\tilde{k}_{x}^{2} +k_{z}^{2} \right)+\tilde{k}_{x}^{2} N^{2} c_{s}^{2} \approx 0.\]

For the solution in form (\ref{22}) we get:
\[\tilde{v}_{x} ,\tilde{v}_{z} \propto \exp \left(z/2H\right)\exp \left(\delta x\right)\exp \left[i\left(\omega t-\tilde{k}_{x} x-k_{z} z\right)\right].\]
Taking into account the substitution of variables made above, the dependence of the AGW amplitude on the wind speed takes the form:
\begin{equation} \label{26}
A\approx A_{0} \left(W_{x} /W_{x0} \right)\exp \left(\delta x\right),
\end{equation}
where $\delta $ is given by formula (\ref{25}). We have an increase in the amplitude of the waves at $\delta >0$. In a counterflow with increasing speed ($\frac{\partial W_{x} }{\partial x} <0$), the amplitude of the waves increases if the expression in the square brackets in (\ref{25}) is positive. The first term in (\ref{25}), which is the largest in magnitude, is always positive for the gravitational branch of the AGW spectrum, since the inequality $\tau ^{2} >1$ holds. The second term may change its sign, but it makes a minor contribution to the amplitude change only at high wind speeds. The third term is positive at $\tau ^{2} >2$. At shorter periods (higher frequencies) it changes the sign, i.e. it actually inhibits amplitude growth of the AGW at high frequencies.

For sufficiently long periods, in comparison with the BV period ($\tau ^{2} >>1$), it is possible to write down the following approximate dependence of the AGW amplitudes on the wind:
$$A_{1} \approx A_{0} \left(W_{x} /W_{x0} \right)\exp \left[-x\frac{\partial W_{x} }{\partial x} \left(\frac{1}{U_{x} } +\frac{\gamma }{4c_{s}^{2} } \left(U_{x} +W_{x} \right)\right)\right].$$
The exponential part (due to the inertia forces) shows the increase in the amplitude of the waves in the headwind ($\frac{\partial W_{x} }{\partial x} <0$) and its attenuation in the downwind ($\frac{\partial W_{x} }{\partial x} >0$).

The amplitude change in the downwind and headwind is illustrated in Fig. 2a, b for three values of fixed frequencies $\omega _{0} $ and the initial value of the wavelength $\lambda _{xo} $. For the case of a downwind, the frequencies are chosen close to the BV frequency, since they quickly decrease in the increasing tailwind. For the headwind, on the contrary, the frequencies $\omega _{0} $ are chosen low so that they do not exceed the BV frequency at sufficiently large values of the wind speed. The initial wavelengths are chosen as follows: for a tailwind $\lambda _{xo} =230$km (increases with increasing wind speed), for a headwind $\lambda _{xo} =1500$km (decreases with increasing wind speed). A background wind is set for modeling, the speed of which increases linearly from 50 m/s to 500 m/s with a spatial gradient $\frac{\partial W_{x} }{\partial x} \approx 10^{-5} $ s$^{-1}$. This corresponds to the condition of slow changes in wind speed $\lambda _{x} \frac{dW_{x} }{dx} <<W_{x} $ for characteristic scales of AGW in the upper atmosphere from hundreds km to several thousand km.

From Fig. 2a we can show that the dependence of the amplitude on the headwind speed is weakly dependent on the initial frequency $\omega _{0} $. Comparison of Fig. 2a with Fig. 1 shows that the theoretical dependencies at different initial frequencies agree well with the experimental data. As can be seen from Fig. 2b, in the tailwind, a slight initial increase in the amplitude is quickly compensated by an exponential decrease in amplitude. As a result, AGWs are rapidly decreasing if the tailwind increasing.
\begin{figure}
\centering
\includegraphics[scale=0.55]{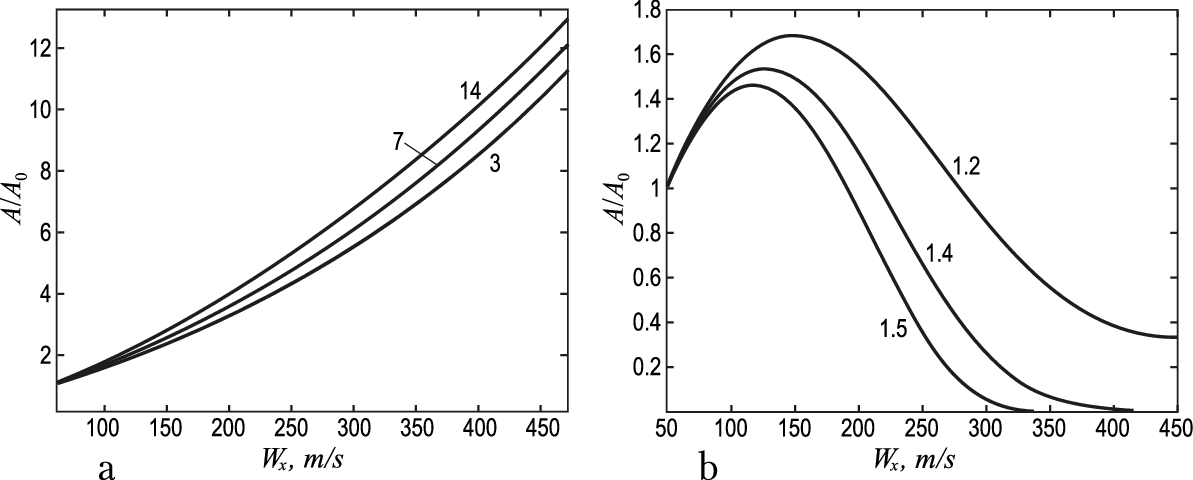}
\caption{Dependence of the AGW amplitude on the wind speed: a) headwind (from top to bottom $N/\omega _{0} =$14, 7, 3 at $\lambda _{xo} =1500$km); b) tailwind (from top to bottom $N/\omega _{0} =$1.2, 1.4, 1.5 at $\lambda _{xo} =230$km).}
\end{figure}

\section{CONCLUSIONS}

In this work, the propagation of AGW in a horizontal wind flow with weak spatial inhomogeneity is investigated. For the analysis, the system of hydrodynamic equations, which takes into account the spatial inhomogeneity of the wind flow, has been used. It is shown that with the help of a special substitution of variables, it is possible to separate the trends of inhomogeneous background parameters of the medium from the effects of inertial forces.

The dispersion equation of AGW in the reference frame of a moving medium is obtained. It coincides with the dispersion equation for a stationary medium, except for the small term. The analytical expression describing the changes in the amplitude of the waves in the medium, moving with a non-uniform speed, is obtained. The dependence of the AGW amplitude consists of two parts: linear and exponential. The linear part is related to the changes in the background parameters of the medium. It does not depend on the direction of wave propagation relative to the flow. The exponential part is caused by the inertial forces and demonstrates the dependence of the amplitudes of AGW on the direction of their propagation (increase and decrease of wave amplitudes in headwind and tailwind, respectively).

The obtained theoretical dependence for AGW amplitudes on a wind speed is in good agreement with the data of satellite observations of these waves in the polar regions of the thermosphere.

The work is supported by the National Research Fund of Ukraine (project 2020.02/0015 "Theoretical and experimental studies of global disturbances of natural and man-made origin in the Earth-atmosphere-ionosphere system"). Fedorenko A.K. and Kryuchkov E.I. thank the State Institution National Antarctic Science Center of the Ministry of Education and Science of Ukraine for the support (contract No.H/02-2023).

\end{document}